\documentclass[prb,twocolumn,aps,showpacs,showkeys,10pt,superscriptaddress]{revtex4}
\usepackage{amsmath}
\usepackage{amssymb}
\usepackage{graphicx}
\usepackage{subfigure}
\usepackage{color}
\usepackage[colorlinks=true,linkcolor=red,urlcolor=blue,citecolor=blue]{hyperref}%specify this as last package
\usepackage{tabularx}

\begin{document}

\title{Electronic structure and magnetic ordering of NiN and Ni$_2$N 
       from first principles}

\author{Abdesalem Houari}
\email[corresponding author:]{abdeslam.houari@univ-bejaia.dz}
\affiliation{Theoretical Physics Laboratory, 
             Department of Physics, 
             University of Bejaia, 
             Bejaia, Algeria}
\author{Samir F.~Matar}
\email[]{s.matar@lgu.edu.lb}
\affiliation{Lebanese German University, LGU, Sahel-Alma Campus, 
             Jounieh, Lebanon}
\author{Volker Eyert}
\email[]{veyert@materialsdesign.com}
\affiliation{Materials Design SARL, 92120 Montrouge, France}

\date{\today} 

\begin{abstract}
The results of first-principles electronic structure calculations for 
the nitrogen-rich nickel nitrides $ {\rm NiN} $ and $ {\rm Ni_2N} $ 
are presented. The calculations are based on density functional theory 
and used the generalized gradient approximation (GGA) as well as the 
GGA$ +U $ approach. The latter turned out to be crucial for a correct 
description of the crystal phase stability and magnetic instabilities 
of both compounds. While for $ {\rm NiN} $ GGA calculations predict 
a non-magnetic ground state with the zincblende structure, GGA$ +U $ 
calculations result in a half-metallic ferromagnet with the rocksalt 
structure in line with indications from the neighboring transition-metal 
nitrides making $ {\rm NiN} $ a possible candidate for spin-filter 
devices. For $ {\rm Ni_2N} $ GGA calculations likewise lead to a 
non-magnetic behavior, which is contrasted with a ferrimagnetic ordering 
obtained from the GGA$ +U $ approach. This ground state results from 
complex three-dimensional exchange interaction via $ \sigma $-type 
and $ \pi $-type overlap of the Ni $ 3d $ orbitals with the N $ 2p $ 
orbitals and may explain the reported sensitivity of the magnetic 
ordering to details of the crystal structure. For both nitrides, 
experimental data are called for to confirm our predictions. 
\end{abstract}

 % Total energy and enthalpy of formation calculations

\maketitle

\section{Introduction}
\label {Intro}

As compared to the huge class of transition-metal oxides the respective 
nitrides have been much less studied due in part to comparably greater 
difficulties to synthesize these systems as a consequence of the high 
stability of the $ {\rm N_2} $ molecule. Nevertheless, a wide spectrum 
of promising electronic, magnetic, and mechanical properties generated 
high interest in the nitrides since long. \cite{toth71} This applies 
especially to the $ 3d $ transition metal nitrides, some of which show 
superconducting phases and magnetic instabilities while others allow 
application, e.g.\ as hard materials, anticorrosion coatings, electrical 
contacts, diffusion barriers, catalysts, sensors, and fuel cells. 
\cite{shimizu97,gajbhiye02} Yet, in 
particular the late transition-metal nitrides long resisted successful 
synthesis, a fact attributed to the increased filling of their 
antibonding $ 3d $-$ 2p $ states. \cite{guillermet94} For this reason, 
the thermodynamic stability, crystal structures as well as the 
electronic and magnetic properties of these latter binaries have been a 
matter of dispute for long and only recently work on these compounds 
has started to measure up with the increasing interest. This holds true 
also for the nickel nitrides. Published phase diagrams include mainly 
three different nitride phases, namely, $ {\rm Ni_8N} $, $ {\rm Ni_4N} $, 
and $ {\rm Ni_3N} $. \cite{guillermet91,neklyudov04} However, even the 
thermodynamic stabilities of some of these phases were controversely 
discussed for a long time. 

Probably the best studied nickel nitride is $ {\rm Ni_3N} $, which was 
obtained already quite early from nickel metal and ammonia. \cite{juza43} 
From x-ray diffraction, the Ni sublattice was identified as a hexagonal 
close-packed structure with lattice parameters $ a = 2.664 $\,\AA\ and 
$ c = 4.298 $\,\AA. \cite{juza43} The nitrogen atoms intercalate into 
the metal sublattice and occupy the octahedral interstices. As a result, 
the crystal structure of $ {\rm Ni_3N} $ assumes the hexagonal 
$ \varepsilon $-$ {\rm Fe_3N} $ structure with space group 
$ {\rm P6_322} $ and the basal-plane lattice parameter increased by a 
factor $ \sqrt{3} $, i.e.\ to $ a \approx 4.62 $\,\AA. 
\cite{juza43,leineweber00,leineweber01} 
$ {\rm Ni_3N} $ like many other binary $ 3d $ transition-metal nitrides 
thus falls into the broad class of metallic interstitial compounds with 
a metal sublattice, which is characteristic of a simple metal (e.g.\ 
fcc, bcc, or hcp), and nitrogen atoms located at the interstices. 
\cite{leineweber01} Possible disorder of the nitrogen atoms was addressed 
in detail by Leineweber {\em et al.}, who investigated the 
$ \varepsilon $-phase nitrides of Mn, Fe, and Ni using neutron powder 
diffraction. They found an order-disorder transition at 550\,K in 
$ {\rm Mn_3N_{1.17}} $, partial disorder with increasing temperature in 
$ {\rm Fe_3N_{1+x}} $, and complete order in $ {\rm Ni_3N} $. 
\cite{leineweber00,leineweber01,leineweber04} 
The overall trend of this behavior was explained by the decrease of 
lattice parameters along the series, which causes decrease in N-N 
distances. The resulting increase in repulsive interactions between 
the interstitial atoms would then cause their ordering. \cite{leineweber00} 
The hexagonal $ \varepsilon $-$ {\rm Fe_3N} $ structure of $ {\rm Ni_3N} $ 
and the above given lattice parameters were subsequently confirmed by 
several groups using powder and nanocrystalline samples as well as thin 
films obtained from various synthesization processes and characterized 
by x-ray, neutron, and electron diffraction. 
\cite{dorman83,gajbhiye02,leineweber01,leineweber04,neklyudov04,vempaire04,guillaume06,vempaire09a,vempaire09b,lindahl09,popovic09,leineweber12} 
Nevertheless, $ {\rm Ni_3N} $ was also reported to thermally decompose 
at temperatures above 600-680\,K. 
\cite{juza43,maya93,leineweber00,leineweber01} Juza and Sachsze 
also mentioned the metallic bonding in this nickel nitride, which contrasts 
the ionic metal-nitrogen bonding reported for $ {\rm Cu_3N} $. \cite{juza43} 

Only few studies focused on the magnetic properties of $ {\rm Ni_3N} $. 
While Gajbhiye {\em et al.}\ found stable ferromagnetic order below 
$ T_{\rm C} = 634 $\,K, Vempaire {\em et al.}\ as well as Leineweber 
{\em et al.}\ did not observe any indications for a magnetic instability. 
\cite{gajbhiye02,leineweber04,vempaire04,popovic09,vempaire09a,vempaire09b}
This finding was also supported by first principles calculations, which 
attributed the suppression of magnetic order to delocalization of the 
electronic states due to the strong Ni $ 3d $-N $ 2p $ hybridization. 
\cite{vempaire04,vempaire09a,vempaire09b,fang12,imai14}

In contrast to $ {\rm Ni_3N} $, $ {\rm Ni_4N} $ and $ {\rm Ni_8N} $ 
were much less studied since they appeared mainly as intermediate 
phases during the decomposition of the former compound. 
\cite{maya93,gajbhiye02,terao60,nagakura73} Nevertheless, $ {\rm Ni_4N} $ 
was characterized as having a cubic lattice with lattice constant 
$ a = 3.77 $\,\AA. \cite{neklyudov04} From first principles calculations, 
both $ {\rm Ni_4N} $ and $ {\rm Ni_8N} $ were predicted to have non-zero 
magnetization, \cite{kong98,fang12,hemzalova13,imai14,meinert16} similar 
to the neighboring transition-metal nitrides $ {\rm M_4N} $ 
($ {\rm M = Mn, Fe, Co} $), which have been studied by various authors. 
\cite{matar88,mohn92,houari07b,monachesi13,meinert16} 
     
The nickel nitrides with a higher nitrogen concentration were even less 
studied than the Ni-rich nitrides. $ {\rm Ni_2N} $ assumes a simple 
tetragonal structure with lattice parameters $ a = 2.80 $\,\AA\ and 
$ c = 3.66 $\,\AA\ and space group $ {\rm P4/mmm} $. \cite{dorman83} The 
magnetic properties of $ {\rm Ni_2N} $ are still unclear. While 
measurements on thin layers as well as first principles calculations 
using the generalized-gradient approximation (GGA) of density functional 
theory (DFT) did not show any magnetic order, more recent experiments 
indicated spin-glass like behavior and proximity of a magnetic 
instability at increased volume. \cite{vempaire09a,vempaire09b,nishihara14} 

Finally, like CuN nickel mononitride seems not to have been investigated 
experimentally so far. \cite{wang04} However, first principles calculations 
using the local density approximation (LDA) or the GGA showed that NiN assumes 
a metallic, non-magnetic ground state within the zincblende structure like 
the neighboring mononitrides CoN and CuN. \cite{wang04,paduani08} 
In contrast, CrN showed antiferromagnetic ordering in an orthorhombic 
structure. \cite{mavromaras94} 
     
The present work, which is part of a broader set of investigations 
on magnetic transition-metal nitrides,  
\cite{matar07,houari07a,houari07b,houari08,houari10a,houari10b} 
focuses on the thermodynamic stability as well as the electronic and 
magnetic properties of $ {\rm NiN} $ and $ {\rm Ni_2N} $. To this end we 
apply electronic structure calculations as based on density functional 
theory within the generalized-gradient approximation (GGA). However, in 
the course of our work it became obvious that a successful description 
of both compounds requires to go beyond and to apply the GGA$ + U $ 
approach in order to take proper account of the local electronic correlations 
impacting the electronic and magnetic properties. In order to be consistent 
and underline the validity of our approach, we complement the calculations 
for the above N-rich compounds by computations for $ {\rm Ni_3N} $. 

The paper is organized as follows: After describing the theoretical methods 
and computational details in Sec.~\ref{methods}, we present the calculated 
results obtained for all three compounds in Sec.~\ref{results} before 
concluding in Sec.~\ref{conclusion}.

\section{Computational Methods}
\label{methods}

The first principles calculations were based on density functional 
theory \cite{hohenberg64,kohn65} with exchange-correlation effects 
accounted for by the PBE parametrization of the generalized gradient 
approximation (GGA) as proposed by Perdew, Burke, and Ernzerhof.  
\cite{perdew96} Motivated by a large amount of studies on the 
paradigmatic antiferromagnetic insulator NiO, for which taking 
account of local electronic correlations via a GGA$ +U $ treatment 
turned out to be crucial, we complemented the GGA calculations by 
calculations using the GGA$ +U $ method. 
\cite{anisimov91,liechtenstein95} Following previous GGA$ +U $ 
investigations on NiO we selected $ U = 6.0 $\,eV and $ J = 0.95 $\,eV 
for the Ni $ 3d $ orbitals. \cite{zhou04,jain11,seth17,ryee17} However, 
to check the sensitivity of the results to the choice of these 
parameters, we also performed some calculations with $ U = 4.0 $\,eV 
and $ J = 0.64 $\,eV. For each set of $ U $ and $ J $ we first 
performed a structure relaxation, which was then followed by a 
calculation of the electronic and magnetic properties. In 
doing so we even distinguished spin-degenerate and spin-polarized 
situations, i.e.\ both structure relaxation and calculation of 
the electronic and magnetic properties were performed separately for 
these two cases. As a consequence, there is an intimate connection 
between $ U $ and $ J $, the crystal structure parameters, and the 
electronic structure. For the double counting correction the fully 
localized limit was used. 
\cite{anisimov93,czyzyk94,solovyev94,liechtenstein95}  

Two complementary first principles methods were employed: 
In a first step, the Vienna Ab initio Simulation Program (VASP) as 
implemented in the MedeA$ \textsuperscript{\textregistered} $ 
computational environment of Materials Design was used to perform 
total-energy and force calculations aiming at a relaxation of the 
structures. \cite{vasp,medea} The single-particle equations were solved 
using the projector-augmented wave (PAW) method with a plane-wave basis 
and a cutoff of 520 eV. \cite{paw,vasppaw} All calculations were 
converged until the total-energy difference between two consecutive 
iterations was lower than $ 10^{-5} $\,eV and the forces on the atoms 
were converged to 0.02 eV/\AA. The Brillouin zone of $ {\rm NiN} $ was 
sampled using a Monkhorst-Pack mesh with a spacing below $ 0.2 $\,\AA$^{-1}$ 
leading to $ 15 \times 15 \times 15 $ {\bf k}-points for the rocksalt 
structure, $ 13 \times 13 \times 13 $ {\bf k}-points for the zincblende 
and cesium chloride structures, and $ 13 \times 13 \times 7 $ 
{\bf k}-points for the wurtzite structure. \cite{monkhorst76} For 
$ {\rm Ni_2N} $ and $ {\rm Ni_3N} $ this spacing led to 
$ 13 \times 13 \times 9 $ and $ 9 \times 9 \times 9 $ {\bf k}-points, 
respectively. 

Once the equilibrium structures were known, analysis of the electronic 
structure, magnetic ordering, and chemical bonding was carried out 
using the full-potential augmented spherical wave (ASW) method in its 
scalar-relativistic implementation. \cite{eyert00,eyert13} 
In the ASW method, the wave function is expanded in atom-centered
augmented spherical waves, which are Hankel functions and numerical
solutions of Schr\"odinger's equation, respectively, outside and inside
the so-called augmentation spheres. In order to optimize the basis set 
and enhance the variational freedom,
additional augmented spherical waves were placed at carefully selected
interstitial sites. The choice of these sites as well as the augmentation
radii were automatically determined using the sphere-geometry optimization
algorithm. \cite{eyert98} Self-consistency was achieved by a highly efficient
algorithm for convergence acceleration \cite{eyert96} until the variation 
of the atomic charges was smaller than $ 10^{-8} $ electrons and the 
variation of the total energy was smaller than $ 10^{-8} $ Ryd. Brillouin 
zone integrations were performed using the linear tetrahedron method 
\cite{bloechl94} with up to 
$ 26 \times 26 \times 26 $ {\bf k}-points for the rocksalt structure, 
$ 24 \times 24 \times 24 $ {\bf k}-points for the zincblende and cesium 
chloride structures, 
and $ 23 \times 23 \times 12 $ {\bf k}-points for the wurtzite structure 
of $ {\rm NiN} $. In contrast, for $ {\rm Ni_2N} $ and $ {\rm Ni_3N} $ 
up to $ 22 \times 22 \times 17 $ and $ 16 \times 16 \times 15 $ 
{\bf k}-points, respectively, were used. Increasing the 
number of {\bf k}-points in several steps up to the just mentioned 
maximum settings allowed to control convergence of the results with 
respect to the density of the Brillouin-zone mesh. 

%In the present work, a new full-potential version of the ASW method was 
%employed. \cite{eyert13} In this version, the electron density and 
%related quantities are given by spherical harmonics expansions inside 
%the muffin-tin spheres. In the remaining interstitial region, a 
%representation in terms of atom-centered Hankel functions is used.  
%\cite{methfessel88} However, in contrast to previous related 
%implementations, no so-called multiple-$\kappa$ basis set is needed, 
%rendering the method computationally nearly as efficient as the 
%original ASW scheme.

\section{Results and Discussions}
\label{results}
   
\subsection{$ {\rm \bf Ni_3N} $}
\label{sect:ni3n}

In order to obtain a first impression of the thermodynamic stability 
and electronic structure of the nickel nitrides and to lay ground for 
the subsequent discussion of the nitrogen-rich class members we 
start out considering $ {\rm Ni_3N} $. From structure relaxations using 
VASP together with both GGA and GGA$ +U $ lattice parameters of 
$ a = 4.617 $\,\AA, $ c = 4.306 $\,\AA\ and $ a = 4.512 $\,\AA, 
$ c = 4.193 $\,\AA, respectively, were obtained. As expected, the 
GGA$ +U $ results are 
slightly smaller than the GGA values due to the stronger localization of 
the Ni $ 3d $ orbitals. The internal parameter fixing the position of the 
Ni atoms was calculated as $ x_M = 0.330 $ and $ x_M = 0.333 $ using 
the GGA and GGA$ +U $ method, respectively. Both sets are close to 
the experimental data of $ a = 4.622 $\,\AA, $ c = 4.306 $\,\AA, and 
$ x_M = 0.3279 $. \cite{leineweber01} The enthalpy of formation as compared 
to the elements in their standard state (ferromagnetic face-centered cubic 
Ni and $ {\rm N_2} $ molecule) was calculated to $ +0.10 $\,eV and 
$ -0.08 $\,eV per formula unit from GGA and GGA$ +U $, respectively. 
Hence, while both values are close to each other, the negative value 
obtained from the GGA+U method confirms the necessity to go beyond the 
GGA. Finally, we point out that neither of these approaches led to 
magnetic order, in agreement with the experimental findings. 

\begin{figure}[htb]
\includegraphics[width=\columnwidth]{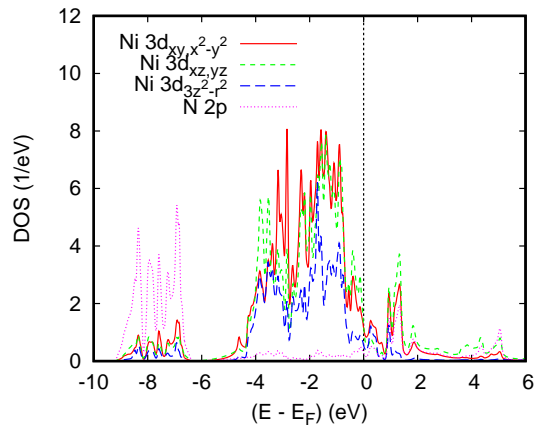}
\caption{Partial densities of states of $ {\rm Ni_3N} $ as arising from 
         the GGA$ +U $ calculations. In all figures a Gaussian broadening 
         of 50\,meV was used.} 
\label{fig:ni3n-gga+u-dos}
\end{figure}

The electronic densities of states as obtained from spin-degenerate ASW 
calculations using the GGA$ +U $ approach, which turned out to be very 
similar to those resulting from the GGA calculations, are displayed in 
Fig.~\ref{fig:ni3n-gga+u-dos}. 
We distinguish the in-plane Ni $ 3d_{xy, x^2-y^2} $, out-of-plane 
Ni $ 3d_{3z^2-r^2} $, and the mixed Ni $ 3d_{xz, yz} $ orbitals. 
Since the out-of-plane orbitals mediate weak overlap perpendicular 
to the basal plane of the hexagonal unit cell with like orbitals 
of Ni sites neighboring along the $ c $ axis, their partial densities 
of states are essentially confined to the central energy interval 
from $ -5 $\,eV to the Fermi energy, where N $ 2p $ contributions 
are small. In contrast, the in-plane and mixed orbitals, in addition 
to dominating in the central energy interval, also contribute to the 
energy intervals from about $ -9.5 $\,eV to $ -6.5 $\,eV and from 
the Fermi energy to about $ +2 $\,eV, where they complement the strong 
N $ 2p $ partial densities of states. This complementarity results 
from the strong spatial overlap of these orbitals, which leads to 
bonding and antibonding manifolds in the lower and upper part of 
the spectrum.

\subsection{$ {\rm \bf Ni_2N} $}
\label{sect:ni2n}

\begin{figure}[htb]
\includegraphics[width=0.64\columnwidth]{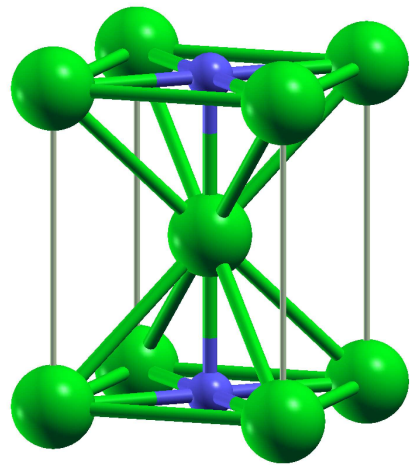}
\caption{Crystal structure of $ {\rm Ni_2N} $. Nickel and nitrogen atoms 
         are given in green and blue, respectively.}
\label{fig:ni2n-gga+u-str}
\end{figure}

According to recent measurements, $ {\rm Ni_2N} $ assumes a simple 
tetragonal structure with space group $ {\rm P4/mmm} $ and lattice 
parameters $ a = 2.815 $\,\AA\ and $ c = 3.665 $\,\AA. \cite{nishihara14,ma17} 
The structure is displayed in Fig.~\ref{fig:ni2n-gga+u-str}. 
The Ni atoms occupy the Wyckoff positions (1a) and (1d) and thus 
are located at the corner (labeled Ni1 below) and the center (Ni2), 
respectively, of the unit cell, whereas the nitrogen atoms occupy 
the Wyckoff position (1c) at the center of the basal plane. Hence, 
the two Ni atoms are not equivalent by symmetry but would be and form 
a body-centered tetragonal lattice were it not for the nitrogen atoms. 
As well known, in the tetragonal crystal system the body-centered 
lattice is identical to a face-centered lattice with the conventional-cell 
vectors rotated by $ 45^{\circ} $ about the tetragonal $ c $ axis, 
the lattice parameter $ a $ increased by a factor of $ \sqrt{2} $, 
and the $ c/a $ ratio reduced by the same factor. \cite{ashcroft} In 
the present case, this would give rise to a face-centered tetragonal 
lattice with one nickel atom per primitive cell and lattice parameters 
$ a = 3.981 $\,\AA\ and $ c = 3.665 $\,\AA, i.e.\ $ c/a = 0.921 $. 
These numbers have to be contrasted with the lattice parameter 
$ a = 3.524 $\,\AA\ found for face-centered cubic Ni. Starting from 
that lattice, the nitrogen atoms are inserted into part of the octahedral 
voids, specifically at the centers of eight out of twelve edges of the 
cubic cell and, hence, give rise to a planar arrangement. As a 
consequence, strong elongation by about 13\% of the corresponding 
in-plane lattice parameters of the original face-centered cubic lattice 
of elemental Ni is found, whereas elongation along the third axis 
amounts to only 4\%. Obviously, even without the insertion of nitrogen 
such geometric changes would have a strong impact on the ferromagnetic 
ordering. In particular, the lattice expansion would cause narrowing of 
the electronic bands, hence, overall increase of the density of states, 
and via the Stoner criterion strengthening of the ferromagnetic order. 
Indeed, as calculations for elemental nickel within the face-centered 
cubic and tetragonal lattices reveal, the lattice expansion leads to 
increase of the ferromagnetic moment from $ 0.63 \mu_B $ to 
$ 0.82 \mu_B $ per atom. 

As was also mentioned above, measurements on thin films gave no hint at 
long-range magnetic order. \cite{vempaire09a,vempaire09b,linnik13} 
While this finding was supported by first principles calculations as 
based on the GGA, \cite{vempaire09a,vempaire09b,nishihara14} 
magnetization measurements on bulk samples revealed spin-glass like 
behavior and initiated calculations, which showed onset of ferromagnetic 
order for an isotropically expanded lattice. \cite{nishihara14} 

Again, in a first step we performed structure relaxations using VASP 
together with both GGA and GGA$ +U $. From the former, lattice parameters 
of $ a = 2.815 $\,\AA\ and $ c = 3.644 $\,\AA\ were obtained. In contrast, 
the latter led to $ a = 2.817 $\,\AA\ and $ c = 3.589 $\,\AA, which as 
expected are smaller than the GGA values due to the stronger localization 
of the Ni $ 3d $ orbitals. Both results are in very good agreement with the 
measured values. The enthalpy of formation as compared to the elements in 
their standard state was calculated as $ +0.46 $\,eV and $ +0.03 $\,eV per 
formula unit from GGA and GGA$ +U $, respectively. Again, the difference 
between these values underlines the necessity to use the GGA$ +U $ approach, 
while the fact that the latter value is still slightly positive reflects 
the conflicting reports about the ground state of this material. Finally, 
we point out that while from the GGA calculations no magnetic order could 
be obtained, the GGA$ +U $ calculations gave rise to a ferrimagnetic ground 
state. Our results are thus in line with the indications of a magnetic 
instability found by Nishihara and coworkers. \cite{nishihara14} The 
ferrimagnetic ordering found from the GGA$ +U $ calculations also explains 
the larger difference between the GGA and GGA$ +U $ values of the enthalpy  
of formation as compared to the corresponding results for $ {\rm Ni_3N} $.  

\begin{figure}[htb]
\includegraphics[width=\columnwidth]{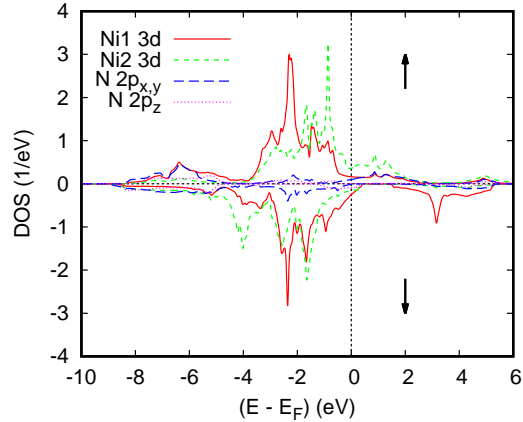}
\caption{Partial densities of states of $ {\rm Ni_2N} $ as arising from 
         the GGA$ +U $ calculations.} 
\label{fig:ni2n-gga+u-dos}
\end{figure}

\begin{figure}[htb]
\includegraphics[width=\columnwidth]{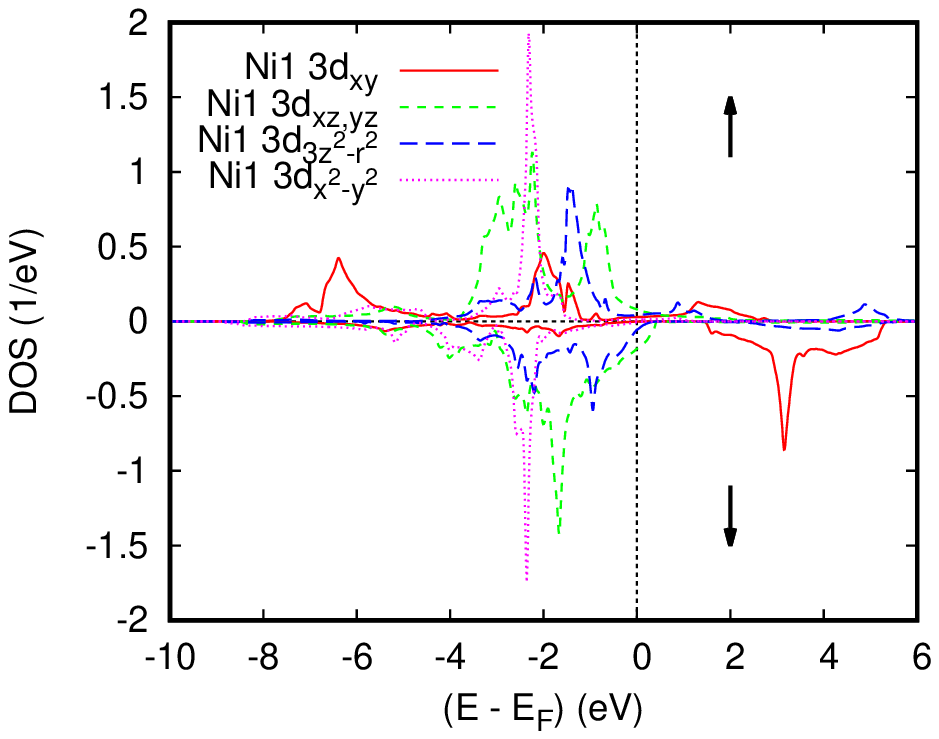}
\caption{Partial Ni1 densities of states of $ {\rm Ni_2N} $ as arising from 
         the GGA$ +U $ calculations.} 
\label{fig:ni2n-gga+u-dos-Ni1}
\end{figure}

\begin{figure}[htb]
\includegraphics[width=\columnwidth]{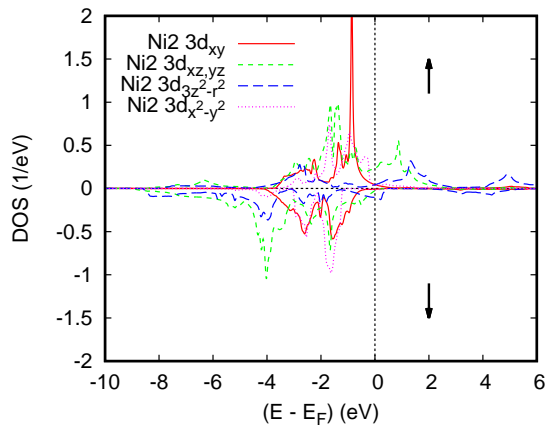}
\caption{Partial Ni2 densities of states of $ {\rm Ni_2N} $ as arising from 
         the GGA$ +U $ calculations.} 
\label{fig:ni2n-gga+u-dos-Ni2}
\end{figure}

The electronic densities of states as obtained from spin-polarized ASW 
calculations using the GGA$ +U $ approach are displayed in 
Figs.~\ref{fig:ni2n-gga+u-dos}, 
\ref{fig:ni2n-gga+u-dos-Ni1}, 
and \ref{fig:ni2n-gga+u-dos-Ni2}. 
In Fig.~\ref{fig:ni2n-gga+u-dos}, we distinguish the partial densities 
of states due to the N $ 2p_{x,y} $ and $ 2p_z $ orbitals, which form 
$ \sigma $-type bonds with the Ni1 $ 3d_{xy} $ and Ni2 $ 3d_{3z^2-r^2} $ 
orbitals, respectively. In case of Ni1 this overlap gives rise to 
distinct spin-up bonding and spin-down antibonding peaks at about 
$ -6.5 $ and $ +3 $\,eV, respectively, which are also clearly observed in 
Fig.~\ref{fig:ni2n-gga+u-dos-Ni1}. In addition, it causes antiparallel 
alignment of the local magnetic moments of $ +0.50 \mu_B $ and 
$ -0.26 \mu_B $, which are carried exclusively by the Ni1 $ 3d_{xy} $ 
and the N $ 2p_{x,y} $ orbitals, respectively. In contrast, the local 
magnetic moments of $ -0.97 \mu_B $ at the Ni2 sites are due to the 
$ 3d_{3z^2-r^2} $, $ 3d_{xz} $, and $ 3d_{yz} $ states but do not 
polarize the N $ 2p_z $ states. Thus, a rather complex magnetic behavior 
is observed, where the magnetic moments at the Ni1 sites interact only 
within the tetragonal basal plane via $ \sigma $-type overlap with the 
in-plane N $ p_{x,y} $ orbitals. In contrast, the magnetic moments at 
the Ni2 sites are subject to two different kinds of exchange interactions, 
namely, via $ \sigma $-type overlap of the Ni2 $ 3d_{3z^2-r^2} $ with 
the unpolarized N $ 2p_z $ states and via $ \pi $-type overlap of the 
Ni2 $ 3d_{xz,yz} $ and the polarized N $ 2p_{x,y} $ orbitals. Finally, 
we observe overlap of like-spin $ 3d_{xz,yz} $ orbitals of both Ni 
sites. While the Ni1 $ 3d_{xz,yz} $ orbitals do not carry any magnetic 
moment they could nevertheless give rise to ferromagnetic superexchange 
interaction of neighboring Ni2 sites both within the Ni2 planes and across. 
Taken together, the just mentioned different kinds of exchange interactions 
establish long-range three-dimensional magnetic order. Note that this 
complex magnetic behavior is a mere consequence of the insertion of 
nitrogen since face-centered tetragonal elemental Ni would show strong 
ferromagnetism as pointed out above. Finally, since the interplay of 
the out-of-plane $ \sigma $-type and $ \pi $-type exchange interactions 
may be rather sensitive to the value of the tetragonal $ c $ axis, thin 
films are likely to show different magnetic properties than bulk samples 
as was indeed observed. In addition, this sensitivity may be responsible 
for the observed spin-glass like behavior. Of course, it would be exciting 
to have our results corroborated by new experimental data.

\subsection{$ {\rm \bf NiN} $}
\label{sect:nin}

As for the other two compounds, in a first step the structural stability 
of NiN was probed by performing spin-degenerate GGA structure relaxations. 
Motivated by conflicting theoretical and experimental results regarding 
the relative stability of the zincblende and rocksalt structures of the 
transition-metal nitrides, \cite{shimizu97,wang04,hong05,miao07,chan08} 
we took different crystal structures into account, namely, the rocksalt 
(RS), zincblende (ZB), cesium chloride (CC), and wurtzite (WZ) structure. 
In the rocksalt and cesium chloride structures both atoms are octahedrally 
coordinated by the respective other species, whereas in the zincblende 
and wurtzite structures the coordination by the respective other species 
is tetrahedral. The results of the structure relaxations are summarized 
in Tab.~\ref{tab:structures-pbe}. 
\begin{table}[htb]
\caption{Strutural parameters and enthalpies of formation as resulting from 
         GGA calculations for NiN in the rocksalt (RS), cesium chloride (CC), 
         zincblende (ZB), and wurtzite (WZ) structures.}
\begin{ruledtabular}
\begin{tabular}{ccccc}
                  &    RS   &   ZB   &   CC   &   WZ   \\ 
\hline
$ a $/\AA         &   4.063 &  4.321 &  2.542 &  3.017 \\ 
$ c $/\AA         &         &        &        &  5.129 \\
$ u $             &         &        &        & 0.3745 \\
$ E_{coh} $/eV    &   +1.25 &  +0.86 &  +2.20 &  +0.90 \\
%$ \Delta E $/mRyd &    28.4 &    0.0 &   98.6 &    2.6 \\ 
\end{tabular}
\label{tab:structures-pbe} 
\end{ruledtabular}
\end{table}
Again, the enthalpies of formation are calculated with respect to 
the total energies of the elements in their standard states. Note 
the trend of the enthalpies of formation to increase on going from 
$ {\rm Ni_3N} $ to $ {\rm Ni_2N} $ and finally to $ {\rm NiN} $, 
which nicely explains the observed higher thermodynamic stability of 
$ {\rm Ni_3N} $ as compared to the N-rich nitrides. In particular, 
from the GGA calculations for NiN all structures listed in 
Tab.~\ref{tab:structures-pbe} are found thermodynamically highly 
unstable. Nevertheless, according to Tab.~\ref{tab:structures-pbe}, 
NiN has the lowest total energy in the zinblende structure in 
agreement with previous calculations. \cite{wang04,paduani08} In 
contrast, the wurtzite structure is found at a slightly elevated 
energy, whereas the other two structures are much higher in energy. 
Unfortunately, no experimental structure determination is available, 
which our results could be compared to. However, since the neighboring 
transition-metal nitrides are found in the rocksalt structure, the 
above finding of the zincblende structure as the most stable one 
leaves some doubt. In passing, we mention that additional spin-polarized 
ferromagnetic calculations led to vanishing magnetic moments for all 
structures. 

\begin{figure}[htb]
\includegraphics[width=\columnwidth]{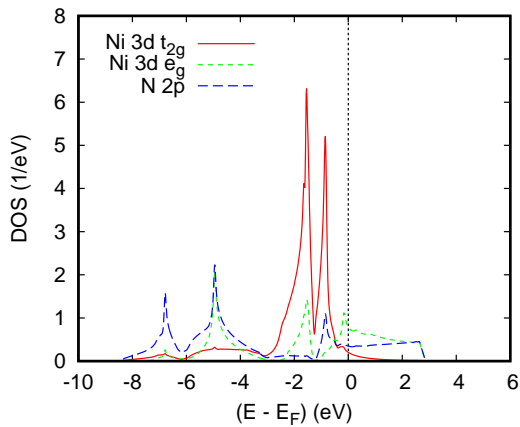}
\caption{Partial densities of states of NiN in the rocksalt structure 
         as arising from the GGA calculations.} 
\label{fig:nin-rs-gga-dos}
\end{figure}

\begin{figure}[htb]
\includegraphics[width=\columnwidth]{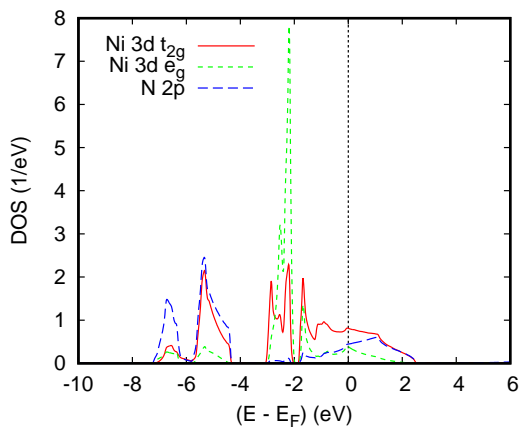}
\caption{Partial densities of states of NiN in the zincblende structure 
         as arising from the GGA calculations.} 
\label{fig:nin-zb-gga-dos}
\end{figure}

Calculated electronic densities of states as obtained from the ASW 
method for the rocksalt and zincblende structure, respectively, are 
displayed in Figs.~\ref{fig:nin-rs-gga-dos}
and \ref{fig:nin-zb-gga-dos}. 
In the rocksalt structure, octahedral arrangement of the Ni atoms by 
nitrogen causes strong $ \sigma $-type overlap of the Ni $ 3d $ $ e_g $ 
states with the N $ 2p $ orbitals, which leads to bonding and antibonding 
manifolds between about $ -8.5 $\,eV and $ -3 $\,eV and above $ -0.5 $\,eV, 
respectively, while the rather non-bonding Ni $ 3d $ $ t_{2g} $ states 
form sharp peaks in between. The situation is quite similar in the 
zincblende structure except for the exchanged roles of the $ e_g $ 
and $ t_{2g} $ orbitals due to the tetrahedral rather than octahedral 
arrangement. For the same reason, the partial densities of states 
obtained for the cesium chloride and wurtzite structure are quite similar 
to those of the rocksalt and zincblende structure, respectively.  

Motivated by the controversial discussion regarding the ground state 
structure and the importance of taking into account local electronic 
correlations for a correct description of the electronic and magnetic 
properties of NiO, we complemented as before the above calculations 
by GGA$ +U $ calculations for NiN. Again, all candidate structures 
were initially relaxed with the results summarized in 
Tab.~\ref{tab:structures-pbepu}.  
\begin{table}[htb]
\caption{Strutural parameters, enthalpies of formation, and local magnetic 
         moment at the Ni site as resulting from GGA$ +U $ calculations 
         for NiN in the rocksalt (RS), cesium chloride (CC), zincblende 
         (ZB), and wurtzite (WZ) structures.}
\begin{ruledtabular}
\begin{tabular}{ccccc}
                  &    RS   &   ZB   &   CC   &   WZ   \\ 
\hline
$ a $/\AA         &   4.115 &  4.241 &  2.496 &  2.957 \\ 
$ c $/\AA         &         &        &        &  5.049 \\
$ u $             &         &        &        & 0.3742 \\
$ E_{coh} $/eV    &   +0.52 &  +0.99 &  +2.22 &  +1.03 \\
%$ \Delta E $/mRyd &   -34.8 &    0.0 &   90.0 &    2.7 \\ 
$ m_{\rm Ni}/\mu_B $       &     1.3 &    0.0 &    0.0 &    0.0 \\ 
\end{tabular}
\label{tab:structures-pbepu} 
\end{ruledtabular}
\end{table}
Obviously, the changes on including local electronic correlations are 
threefold. While the zincblende, cesium chloride, and wurtzite structure 
experience a volume decrease due to the stronger localization of the 
Ni $ 3d $ states, the rocksalt structure shows a volume increase. 
The latter is related to the emergence of long-range ferromagnetic order 
arising from the local moments at the Ni sites. At the same time, the 
magnetic order induces a stabilization of the rocksalt structure, which 
becomes lowest in energy. However, despite this considerable downshift, 
even within the GGA+U approach the enthalpies of formation are still all 
positive, which explains the fact that the late 3d transition-metal 
nitrides have not yet been investigated experimentally. Yet, as already 
mentioned for $ {\rm Ni_2N} $, preparation of samples with reduced 
dimensionality or of nanostructures may still be possible. 

These results perfectly complement those obtained for MnN, where 
GGA/LDA calculations led to a spin-degenerate zincblende structure as 
the ground state, \cite{hong05,miao07} in contradiction with the 
antiferromagnetic distorted rocksalt structure observed experimentally, 
\cite{suzuki00} and the discrepancy had been resolved by GGA$ +U $ 
calculations, which reproduced the experimental findings. \cite{chan08}  

The electronic densities of states as arising from spin-degenerate 
GGA$ +U $ calculations for the so relaxed rocksalt and zincblende 
structures closely resemble those shown in Figs.~\ref{fig:nin-rs-gga-dos} 
and \ref{fig:nin-zb-gga-dos} except for an overall downshift of the 
$ \pi $-type non-bonding states by about 1\,eV relative to the 
bonding and antibonding states resulting from the $ \sigma $-type 
overlap of the Ni $ 3d $ and N $ 2p $ states. The same holds true for 
the densities of states obtained from GGA$ +U $ calculations for the 
cesium chloride and wurtzite structures. 

\begin{figure}[htb]
\includegraphics[width=\columnwidth]{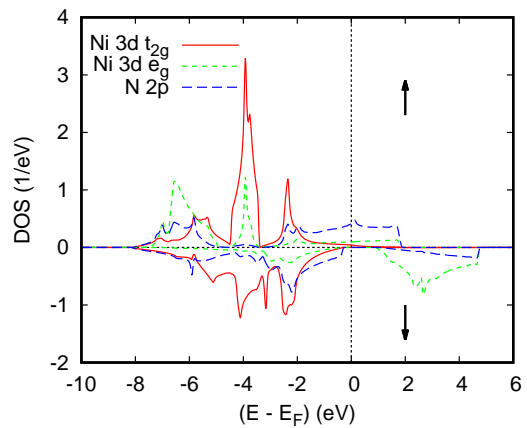}
\caption{Partial densities of states of NiN in the rocksalt structure 
         as arising from the GGA$ +U $ calculations.} 
\label{fig:nin-rs-gga+u-fe-dos}
\end{figure}

The partial densities of states found from spin-polarized ferromagnetic 
calculations for the rocksalt structure are displayed in 
Fig.~\ref{fig:nin-rs-gga+u-fe-dos}.
There, we recognize for both spin channels the sequence of bonding and 
antibonding Ni $ 3d $ $ e_g $ and N $ 2p $ states embracing the weakly 
$ \pi $-bonding Ni $ 3d $ $ t_{2g} $ levels as already observed in 
Fig.~\ref{fig:nin-rs-gga-dos}. As a consequence, the latter states are 
fully occupied and the local magnetic moment of $ 1.3 \mu_B $ is carried 
exclusively by the $ e_g $ states. The imbalance between the spin-up and 
spin-down $ e_g $ electrons is connected with the fact that in the 
spin-majority channel the $ \sigma $-type bonding and antibonding states 
are dominated by the Ni $ 3d $ $ e_g $ and N $ 2p $ states, respectively, 
whereas in the spin-minority channel this is reversed. As a consequence, 
the nitrogen sites carry a small magnetic moment of $ -0.3 \mu_B $, 
which leads to a total magnetic moment per formula unit of $ 1.0 \mu_B $. 
Finally, crystal-field splitting causes opening of a band gap of almost 
1.2\,eV in the spin-minority spectrum, which leaves NiN as a half-metallic 
ferromagnet much like, e.g.\ $ {\rm CrO_2} $, which belongs to the 
most-investigated materials of this exciting class.  
\cite{matar92,matar94,bisti17} 

As expected, results obtained from additional GGA$ +U $ calculations with 
$ U = 4.0 $\,eV and $ J = 0.64 $\,eV are found between those arising from 
the GGA and the above described GGA$ +U $ calculations. In particular, the 
volume increase on going from GGA to GGA$ +U $ as mentioned above is 
reduced as is the stability of the rocksalt structure as compared to the 
zincblende structure. At the same time, the gap in the spin-minority 
channel is also reduced and shifted slightly below the Fermi energy while 
the local magnetic moments carried by Ni and N are reduced to about 
$ 1.1 \mu_B $ and $ -0.1 \mu_B $, respectively. As a deeper analysis 
revealed, these changes result to a large part from the shrinking of 
the lattice coming with the reduction of the local-correlation parameters. 
In contrast, variation of $ U $ and $ J $ while keeping the lattice 
parameter at the value given in Tab.~\ref{tab:structures-pbepu} preserves 
the half-metallic behavior to much lower values of these parameters. 
Nevertheless, in closing we point out that for $ U = 6.0 $\,eV and 
$ J = 0.95 $\,eV as determined and adopted by most researchers the 
calculations clearly show the half-metallic behavior described above. 

Finally, we mention spin-polarized GGA and GGA$ +U $ calculations 
starting from antiferromagnetic order with a spin-propagation vector 
along either the $ \langle001\rangle $ or the $ \langle111\rangle $ 
direction for the cubic structures and along the $ \langle0001\rangle $ 
direction in case of the wurtzite structure, neither of which converged 
to a stable antiferromagnetic state. In addition, we were not able to 
stabilize antiferromagnetic order in the tetragonally distorted rocksalt 
and zincblende structures, contrary to the antiferromagnetic alignment 
found for CrN and MnN, respectively, with orthorhombically and tetragonally 
distorted rocksalt structure. \cite{mavromaras94,chan08} 
   
\section{Conclusion}
\label{conclusion}

The nickel nitrides $ {\rm Ni_3N} $, $ {\rm Ni_2N} $, and $ {\rm NiN} $ 
have been investigated by means of first principles electronic structure 
calculations as based on density functional theory. Both the generalized 
gradient approximation (GGA) and the GGA$+U$ approach were used. While 
for $ {\rm Ni_3N} $ the results obtained from both schemes are very 
similar, taking into account local electronic correlations via the 
GGA$ +U $ approach turned out critical especially for the correct 
description of the magnetic properties of the N-rich compounds. 

For $ {\rm NiN} $, a structural phase study including the rocksalt, 
zincblende, cesium chloride, and wurtzite structures led to diverse 
results. Whereas from GGA calculations the zinblende structure was 
found most stable with a non-magnetic metallic ground state, GGA$+U$ 
calculations led to a half-metallic ferromagnet crystallizing in the 
rocksalt structure as the lowest-energy state. This result is in line 
with both experimental and theoretical findings for the neighboring 
transition-metal nitrides and calls for experimental confirmation, 
in particular in view of possible exciting applications of this material 
in spin-filtering devices. 

For $ {\rm Ni_2N} $, which crystallizes in a simple tetragonal lattice, 
GGA calculations again found a non-magnetic metallic state contrasting 
experimental observations, whereas from GGA$+U$ calculations 
ferrimagnetic order with antiparallel alignment of the moments at the 
two Ni sites and substantial polarization of the N $ p $ states was 
obtained. Coupling of the Ni magnetic moments is via complex interplay 
of three-dimensional exchange interactions, which fact may explain the 
reported sensitivity of the magnetic order to details of the crystal 
structure. Again, refined experiments are called for to confirm our 
predictions.

\begin{acknowledgments}
A.H. gratefully acknowledges financial support of Bejaia University.
\end{acknowledgments}

\end{document}